# Nonpolar resistive memory switching with all four possible resistive switching modes in amorphous ternary rare-earth LaHoO$_3$ thin films

Yogesh Sharma[1], Shojan P. Pavunny[1], Esteban Fachini[2], James F. Scott[3], and Ram S. Katiyar[1]*

[1]Department of Physics and Institute for Functional Nanomaterials, University of Puerto Rico, San Juan, PR-00936-8377, USA.

[2]General Studies College, University of Puerto Rico, San Juan, PR-00931, USA.

[3]Cavendish Laboratory, Department of Physics, University of Cambridge, CB0 3HE, United Kingdom.

## Abstract

We studied the resistive memory switching in pulsed laser deposited amorphous LaHoO$_3$ (a-LHO) thin films for non-volatile resistive random access memory (RRAM) applications. Nonpolar resistive switching (RS) was achieved in Pt/a-LHO/Pt memory cells with all four possible RS modes (i.e., positive unipolar, positive bipolar, negative unipolar, and negative bipolar) having high R$_{ON}$/R$_{OFF}$ ratios (in the range of ~10$^4$-10$^5$) and non-overlapping switching voltages (set voltage, V$_{ON}$ ~ ± 3.6-4.2 V and reset voltage, V$_{OFF}$ ~ ± 1.3-1.6 V) with a small variation of about ± 5-8%. X-ray photoelectron spectroscopic studies together with temperature dependent switching characteristics revealed the formation of metallic holmium (Ho$^0$) and oxygen vacancies (V$_O$) constituted conductive nanofilaments (CNFs) in the low resistance state (LRS). Detailed analysis of current-voltage characteristics further corroborated the formation of CNFs based on metal-like (Ohmic) conduction in LRS. Simmons-Schottky emission was found to be the dominant charge transport mechanism in the high resistance state.

*Corresponding author: e-mail: rkatiyar@hpcf.upr.edu (R. S. K). Tel.: 787 751 4210. Fax: 787 764 2571.

## I. Introduction

Based on the electrical switching of resistivity between high and low conductive states (logic states '1' and '0') of the insulator in a simple metal/insulator/metal (MIM) stack structure, resistive random access memory (RRAM) possesses the advantages of low power consumption, high speed storage capacity, nondestructive readout, simple design (one-transistor and one resistor 1T-1R or cross-bar architecture), compatibility with CMOS (complementary metal-oxide semiconductor) manufacturing process, and excellent scalability, which has been identified/adopted by ITRS (International Technology Roadmap for Semiconductors)[1] as a highly competitive candidate for the next generation of nonvolatile memories. Over the years many metal oxide, chalcogenide and carbon based materials, such as (i) transition metal oxides (TMOs)[2,3] which are dielectric (ii) perovskite-type complex TMOs[4-8] which are multifunctional: paraelectric, ferroelectric, multiferroic, and magnetic, (iii) large bandgap high-k dielectric materials, eg: $Al_2O_3$[9] and rare-earth-oxides,[10,11] (iv) selinides and tellurides,[2,3] and (v) graphene oxide thin films[12] have been found to exhibit unipolar and bipolar resistive switching characteristics. Very recently, nonpolar resistive memories showing four modes of resistive switching (i.e., positive unipolar, positive bipolar, negative unipolar, and negative bipolar) were reported in silicon carbide and magnesium oxide thin films.[13-15] By "nonpolar resistive switching—NRS" we refer to switching devices having both unipolar and bipolar switching characteristics which do not depend on the polarity of the applied voltages. Such nonpolar switching devices possess the advantages of faster switching speed, better uniformity (features of bipolar switching), and high integration density (a feature normally of unipolar switching) along with those mentioned above, which can potentially facilitate advanced applications for RRAMs.[5]

Ternary rare-earth oxides (e.g., $LaGdO_3$, $LaErO_3$, $LaLuO_3$, $SmGdO_3$) are engendering significant research interest as potential materials for microelectronic applications such as metal-oxide-semiconductor structures in transistors, radio frequency (RF) coupling and bypass capacitors in oscillators and resonator circuits, filter and analog capacitors in analog/mixed-signal (AMS) circuits, decoupling capacitors for microprocessors (MPUs), storage capacitors in dynamic random access memory (DRAM) and logic devices and more importantly metal sandwiched resistive layer in RRAM elements.[16-18] These oxides showed high permittivity (~22-31), large optical band gaps (~5.6-5.8 eV), and good compatibility with silicon, which are promising for CMOS and bipolar (Bi)-CMOS chip applications.[16-18] Lanthanum oxide ($La_2O_3$) based ternary rare-earth oxides of general formula $LaREO_3$ (where RE= rare-earth of smaller ionic radii -- i.e., Dy, Ho, Er, Tm, Yb, and Lu) have added advantages of having higher moisture resistance due to comparatively higher lattice energy.[19] Among rare-earth oxides, $Ho_2O_3$ was observed to show the highest lattice energy, which could further minimize the effect of hygroscopicity and consequently improve the high-k properties of ternary $LaHoO_3$.[19] Lately some of these ternary rare-earth oxides have also attracted considerable attention as promising candidates for next-generation non-volatile RRAMs especially due to their semiconductor process compatibility. The MIM capacitor structures of these oxides showed nonvolatile unipolar resistive switching, where the change in conduction behavior of the insulating ternary rare-earth oxide film was explained by a filamentary (thermo-chemical) model.[20,21] The conductive filament formation in these oxides was found to originate from the electric field-induced localized agglomeration of chemical defects present in pristine films.[20] In ternary $LaLuO_3$ a model of resistive switching (RS) based on correlated barrier hopping (CBH) was proposed, where the change in resistance states of the oxide film was attributed to the change in the separation between oxygen vacancy sites.[22] Recently we reported nonvolatile multilevel unipolar resistive switching in amorphous $SmGdO_3$ (SGO) thin films, where the stable 4-level resistance states of a Pt/SGO/Pt memory device with marginal resistance ratios sufficient for real devices were observed by varying compliance current.[23]

Surprisingly, reports on nonpolar switching (with all four possible RS modes) either in binary or ternary rare-earth oxide compounds are scarce in literature. In this paper we report the NRS behavior observed in the ternary high-k dielectric $LaHoO_3$ (k ~32).[24] A Pt/a-LHO/Pt device based on amorphous $LaHoO_3$ (a-LHO) showed repeatable NRS with high $R_{OFF}/R_{ON}$ resistance ratio (~$10^5$) and excellent retention and endurance performances. The four distinct RS modes are explained based on the formation and rupture of conductive nano-filaments. The current conduction mechanisms in ON and OFF-states of the device corresponding to all four RS modes are discussed.

**II. Experimental details**

Polycrystalline LHO powders were synthesized using high-energy solid state reaction route from a stoichiometric mixture (1:1 molar ratio) of $La_2O_3$ and $Ho_2O_3$ powders. High purity (>99.99%) precursors from Alfa Aesar were prefired at 500 °C in argon atmosphere for about two and half hours to remove water content (moisture) and other volatile impurities. Mechanical ball milling was carried out overnight in methanol medium for the fine mixing of resultant molar mixture of $La_2O_3$ and $Ho_2O_3$ powders. As obtained mixture was dried at 500 °C and calcined in air at 1350 °C for 15 hours using a Carbolite HTF1700 furnace with a heating and cooling rate of 5 °C/min followed by an intermediate grinding and heating step. After the calcination steps, the LHO powder with 4-5 wt% of polyvinyl alcohol (PVA) was pressed in the form of thick pellet (diameter = 24 mm, thickness = 3.2 mm) at a uniaxial pressure of 6 ton. Subsequently, this pellet was sintered at 1500 ºC for 12 hours and a highly dense polycrystalline LHO ceramic target was obtained.

LHO films of about 40-nm thickness were fabricated by pulsed laser deposition (PLD) on commercially available Pt/TiO$_2$/SiO$_2$/Si substrates kept at a fixed temperature of 300 °C and ~30 mTorr of ambient oxygen partial pressure inside the deposition chamber. A KrF excimer laser beam (248nm, 10 Hz) that has passed through an aperture to obtain a homogenous flat-top beam profile was used to ablate the polycrystalline LHO ceramic target at an energy density of ~1.7 J/cm$^2$. The total number of shots from the excimer laser on target was 1500 to grow a 40 nm thick LHO layer on the substrate. Platinum top electrodes of ~70 thickness nm and ~80 μm diameter were dc-magnetron sputtered (power density ~1W/cm$^2$) at room temperature through square metal shadow mask to form Pt/LHO/Pt (MIM capacitor structures) resistive memory devices. These devices were passivated using forming gas annealing treatment (90% N$_2$ + 10% H$_2$) at 450 °C for 20 min in rapid thermal annealing (RTA) chamber to reduce the Pt/LHO interface trap density. Structural properties of the films were studied *in situ* by recording reflection high energy electron diffraction (RHEED) patterns at 24 keV beam energy and at 1.4 A filament current and *ex situ* by x-ray diffraction (XRD) measurements using Rigaku Ultima *III*- x-ray diffractometer operating in Bragg-Brentano geometry. The morphology of LHO films were investigated by atomic force microscopy (AFM) using a multimode Nanoscope V (Veeco-AFM) instrument operating in contact mode and by field-emission scanning electron microscopy employing a JEOL JSM-7500F microscope. Material compositions of as-deposited LHO films were analyzed by x-ray photoelectron spectroscopy (XPS). The switching characteristics and conduction mechanisms of the resulting devices were studied through current-voltage (I-V) measurements using a Keithley 2401 sourcemeter unit. A programmable Joule-Thompson thermal stage system (MMR model # K-20) was used for temperature dependent measurements.

**III. Results and discussion**

Figure 1(a) shows *in situ* RHEED pattern from LHO/Pt/TiO$_2$/SiO$_2$/Si heterostructure. The featureless RHEED pattern revealed amorphous phase formation of LHO layer with higher entropy and weak interatomic interaction and suggests that the deposition temperature of 300 $^o$C is sufficient to inhibit crystalline phase formation. X-ray diffraction patterns were obtained from Pt/TiO$_2$/SiO$_2$/Si substrate, and LHO thin film. (c) Cross-sectional FE-SEM image of the LHO/Pt/TiO$_2$/SiO$_2$/Si heterostructure where the zooming view shows the LHO, Pt, and TiO$_2$ layers, respectively. (d) 3-D AFM image of as-grown LHO film.

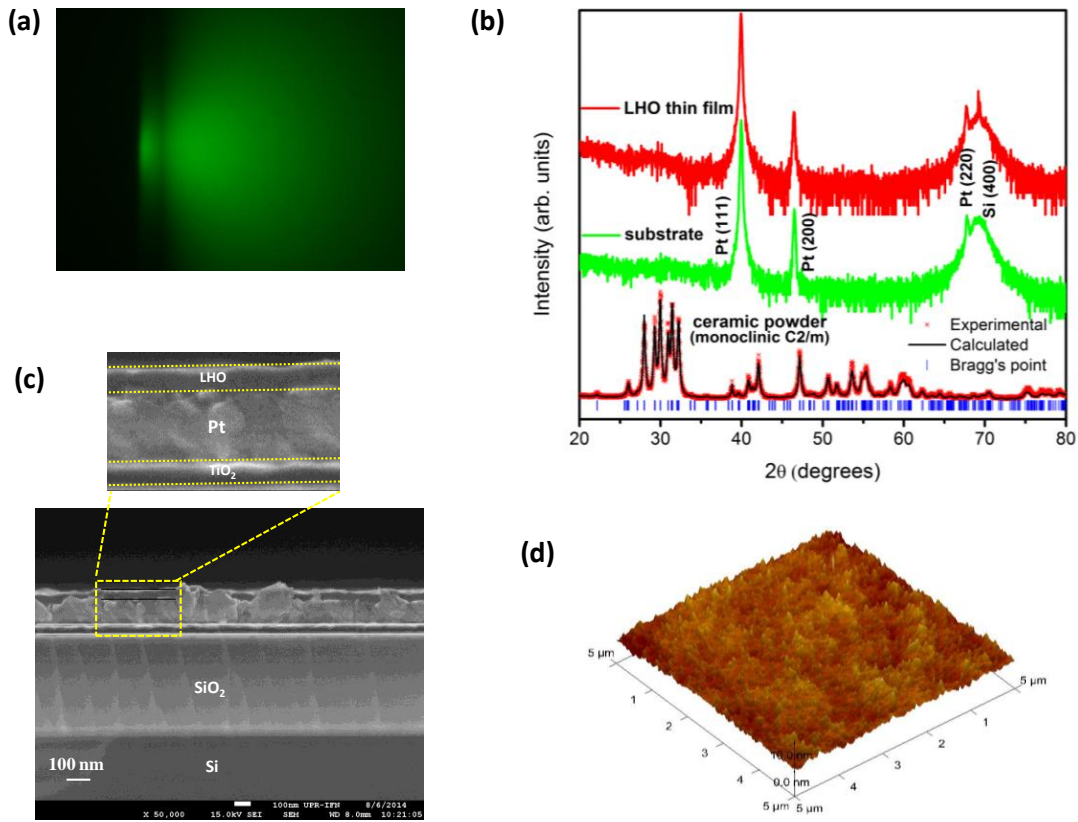

**Figure 1.** (a) The *in situ* RHEED pattern from LHO/Pt/TiO$_2$/SiO$_2$/Si layer during pulsed laser ablation. (b) X-ray diffraction patterns of LHO ceramic powder (used for target preparation),

LHO polycrystalline ceramic powder, substrate, and LHO film, as depicted in Fig. 1(b). As can be seen from Fig. 1(b), all characteristic peaks in the Rietveld least squares minimized powder XRD pattern can be well fitted using centrosymmetric space group $C_{2h}^3$ or $C2/m$ of monoclinic symmetry confirming the single phase formation of bulk LHO.[25,26] According to Fig. 1(b), no Bragg peaks were observed in the XRD pattern of LHO film except those from Pt/TiO$_2$/SiO$_2$/Si substrate which confirm the amorphous phase formation of the film, in agreement with the RHEED results. LHO layer thickness was measured accurately to be ~40 nm from the FE-SEM cross-section image of the LHO/Pt/TiO$_2$/SiO$_2$/Si hetrostructure, as illustrated in Fig. 1(c). Surface topography analysis carried out by AFM measurements demonstrated the homogeneity of the film and the measured root mean square (rms) roughness of the thin layer was found to be ~1.4 nm, as shown in Fig. 1(d).

The illustrative pictures of Pt/a-LHO/Pt/TiO$_2$/SiO$_2$/Si heterostructure and Pt/a-LHO/Pt memory cell are represented in Fig. 2(a). The current-voltage (I-V) characteristics of Pt/a-LHO/Pt memory cell are shown in semi-logarithmic plot in Fig. 2(b)-(f). The Pt/a-LHO/Pt memory cell was found to exhibit NRS with four RS modes: polarity independent unipolar RS (URS) and polarity dependent bipolar RS (BRS) behaviors, under positive and negative DC voltage sweeps with compliance current (I$_{CC}$) of 1 mA. All of the four switching modes were tested on the same cell; the reliability of the memory cell was confirmed by repeated set/reset switching cycles corresponding to each mode without any possible influence of switching history. In the initial state the Pt/a-LHO/Pt memory cell remained in an insulating state with a resistance of ~15 MΩ (read at 0.1 V). An electroforming process was required only for the first switching mode (i.e., positive unipolar) to set the device from its high resistance state (HRS) to a low resistance state (LRS), at a forming voltage of ~6.5 V, as shown in Fig. 2(b).

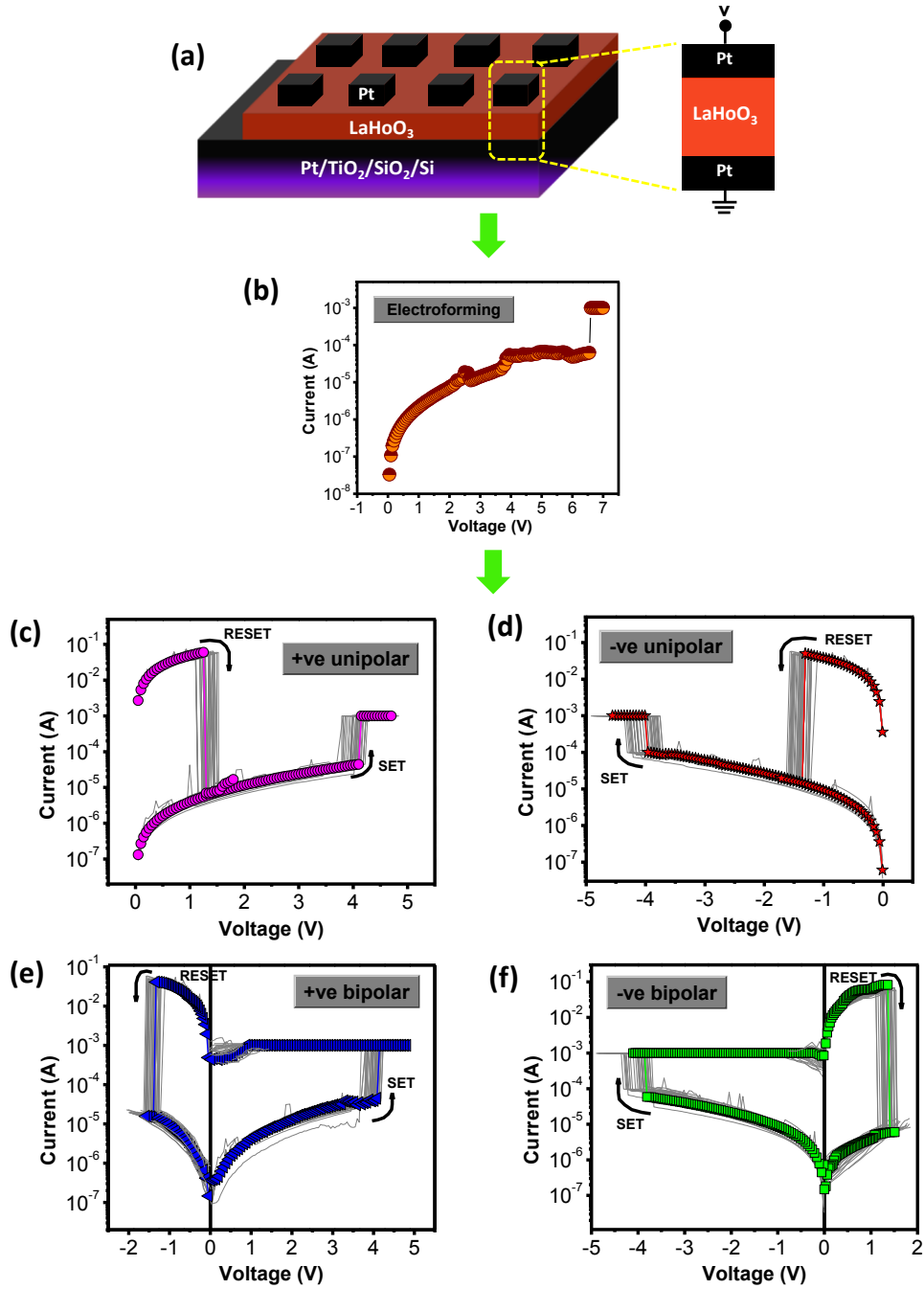

**Figure 2.** (a) The schematic diagrams of Pt/a-LHO/Pt/TiO$_2$/SiO$_2$/Si heterostructure and Pt/a-LHO/Pt MIM memory cell. Typical I-V characteristics of the Pt/a-LHO/Pt memory cell in semi-logarithmic scale: (b) electroforming cycle needed to set the memory cell from initial state and four RS modes: (c) positive URS mode, (d) negative URS mode, (e) voltage polarity dependent positive BRS mode, and (f) negative BRS mode.

The Pt/a-LHO/Pt memory cell showed non-volatile RS behavior in the positive unipolar switching mode, as illustrated in Fig. 2(c). Similar non-volatile RS was observed for the negative unipolar switching mode by applying a negative DC voltage sweep, as shown in Fig. 2(d). After studying these voltage polarity-independent URS modes, we examined the polarity-dependent BRS behavior of the memory cell. As shown in Fig. 2(e) and (f), we found positive as well as negative BRS modes by sweeping DC voltage in either direction to achieve the set/reset processes. Therefore the Pt/a-LHO/Pt device demonstrated NRS with all four possible RS modes: (i) positive unipolar, (ii) negative unipolar, (iii) positive bipolar, and (iv) negative bipolar.[14] More importantly, we noticed that the device can be arbitrarily switched from one switching mode to any of the other switching modes.

The switching mechanism behind the RS phenomenon in the Pt/a-LHO/Pt memory cell was studied by analyzing the temperature dependent switching characteristics of both LRS (ON-state) and HRS (OFF-state). Figure 3(a) and (b) indicate ON and OFF-state resistances as a function of temperature for both positive unipolar and positive bipolar modes. To circumvent high electric field effects, the resistance was measured at a low reading voltage of 0.1 V over the temperature range of 250–460 K. Metallic conduction behavior was observed for the ON-state resistances corresponding to both positive unipolar (+R$_{ON,uni}$) and positive bipolar (+R$_{ON,bi}$) RS modes. The linear fitting of the ON-state resistance versus temperature curves, using the relation; $R_T =$

$R_O[1 + \alpha(T - T_O)]$, where $R_O$ is the resistance at $T_O$, provided the temperature coefficient ($\alpha$) for +$R_{ON,uni}$ and +$R_{ON,bi}$ as 4.1 x$10^{-3}$ K$^{-1}$ and 3.8 x$10^{-3}$ K$^{-1}$, respectively, as shown in Fig. 3(a).

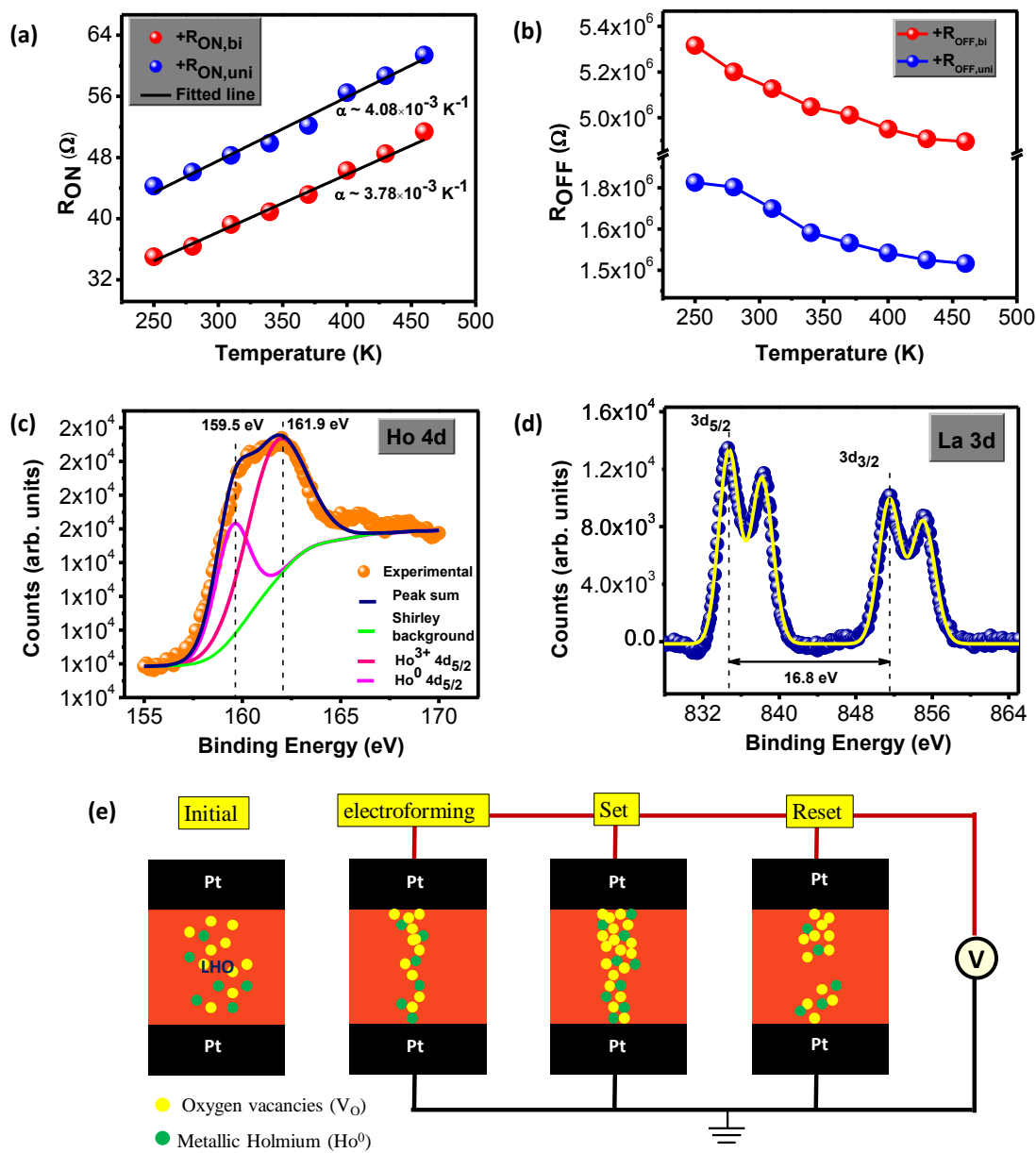

**Figure 3.** Temperature dependence of resistance: (a) ON–state and (b) OFF–state of the device corresponding to positive URS and positive BRS modes. XPS spectra of (c) Ho 4d and (d) La 3d regions of pristine LHO film. Ho 4d$_{5/2}$ peak was deconvoluted into two peaks corresponding to Ho$^{3+}$ and metallic Ho states, respectively. La 3d spectra revealed +3 valence state of La where the two peaks corresponding to 3d$_{3/2}$ and 3d$_{5/2}$ are found to be separated by ~16.8 eV without any reduced bonding signals. (e) Illustrative diagrams for the configurations of oxygen vacancies and metallic Ho in memory cell in initial state and after electroforming, set (ON), and reset (OFF) processes.

The magnitude of these $\alpha$ values obtained is in fairly good agreement with those figures reported for metallic nanowires [i.e. for Cu ~ 2.5 x10$^{-3}$ K$^{-1}$ and for Au ~ 2.3 x10$^{-3}$ K$^{-1}$].[14,24] On the other hand, +R$_{OFF,uni}$ and +R$_{OFF,bi}$ were found to be decreasing with increasing temperature (Fig. 3(b)), as observed in insulators and semiconductors. From these results, we assume that the observed RS behavior in the Pt/a-LHO/Pt memory cell fits to the conductive filamentary model.[9,23,27,28] Such conductive filaments or channels in LHO can be expected due to the alignment of the pre-existing and/or field-induced point defects (i.e., oxygen vacancies and/or metallic rare-earths) under high external electric field during the electroforming process.[23] To validate our assumption, XPS measurements were performed on pristine LHO films to investigate the chemical states of the films. Binding energy corresponding to C 1s peak was utilized to correct the energy shift of La 3d, Ho 4d, and O 1s core levels to account for the x-ray induced insulator charging that otherwise can cause a systematic experimental error. As shown in Fig. 3(c), the XPS analysis of LHO thin film confirmed the existence of metallic Ho, where the Ho 4d$_{5/2}$ peak was deconvoluted into two peaks corresponding to Ho$^{3+}$ (~161.9 eV) and metallic Ho (~159.5 eV) states. We observed that XPS spectra of La did not show any signature of metallic or reduced bonding signal as the two peaks corresponding to La 3d$_{3/2}$ and 3d$_{5/2}$ were found to be located at ~834.4 and ~851.2 eV with a separation of ~16.8 eV, as shown in Fig. 3(d).[21,22] In the LHO film metallic Ho originate from oxygen vacancies which can be described by the following defect Equations (1) and (2) (Kroger-Vink notation):

$$3(O_{O^{2-}}^{2-})^\times = 3(V_{O^{2-}})^{\bullet\bullet} + 6e' + \frac{1}{2}O_2(g) \tag{1}$$

where $(V_{O^{2-}})^{\bullet\bullet}$ denotes an oxygen vacancy. Further, the Ho$^{3+}$-ions in the vicinity of these oxygen vacancies capture the electrons and are reduced to metallic Ho:

$$2\left(Ho^{3+}_{Ho^{3+}}\right)^{\times} + 6e' \rightarrow 2\left(Ho^{0}_{Ho^{3+}}\right)''' \qquad (2)$$

Therefore, the oxygen vacancies in LHO films can cause metallic Ho as reported previously in case of binary rare-earth oxides.[29] Such defects contribute to the formation of conductive filaments or channels during the electroforming process, where the oxygen vacancies and metallic Ho atoms are connected into a channel to form mixed conducting paths, as represented by the illustrative diagram in Fig. 3(e). In the ON-state of the memory cell, electron hopping transport occurred among localized oxygen vacancies and metallic Ho atoms and the flow of current in the device can be expected to be confined within the conductive channels, leading to metallic conduction in ON-state. Based on a filamentary (thermo-chemical) model, the OFF-state of the memory cell can be achieved by the rupture of the conductive channel mainly as a consequence of Joule–heating.[30] The high temperature generated by the Joule-heating thermally activates oxygen migration back to the oxygen vacancy sites and thus oxidizes the metallic Ho atoms present in the conductive channel.[29] Annihilation of oxygen vacancies and oxidation of metallic Ho can thus be attributed to the rupture of filaments, which renders the memory cell into its OFF-state. Therefore, we believe that NRS in the Pt/a-LHO/Pt memory cell can be explained by the localized agglomeration of chemical defects and consequently generation of charged nano-filaments (CFs) that switch the memory cell into ON-states, whereas thermally assisted electromigration of oxygen ions, independent of voltage polarity, ruptures the CF's leading to the memory cell into OFF-states.

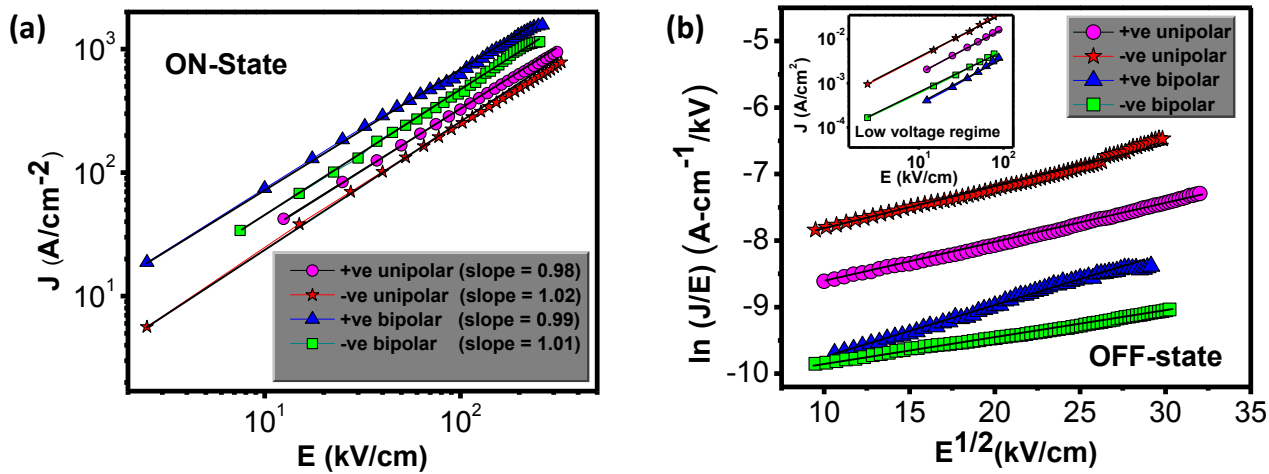

**Figure 4.** (a) ON-state J-E characteristics in double-log plots along with their respective linear fittings corresponding to all four RS modes showing metal like conduction. (b) OFF-state J-E characteristics corresponding to all four RS modes along with the straight line fit using Simmons' modified Schottky emission model [Eq. 3] in high voltage regime (> 0.5 V). Inset shows the OFF-state linear J-E characteristics at low voltage regime (< 0.5 V) validating Simmons' modified Schottky emission conduction mechanism.

To model the conduction mechanism in ON and OFF-states of the memory cell for all four RS modes, the current density (J) versus applied electric field (E) curves were replotted in log-log scale, as illustrated in Fig. 4. In Fig. 4(a), J-E characteristics for all RS modes corresponding to ON-state were found to follow straight line fit with a slope of ~1 indicating metallic (Ohmic) conduction, which supports the model of filamentary conduction in the ON-states of the device.[31] The OFF-state J-E characteristics for all four RS modes were also found to be linear in low voltage regime, whereas at higher bias voltages they showed nonlinear behavior. Such a conduction behavior, especially in oxide films, can be explained by Simmons' modified Schottky Emission mechanism, formulated as:[32-34]

$$J(E) = \alpha T^{3/2} E \mu \left(\frac{m^*}{m_0}\right)^{3/2} exp\left[-\left(\varphi_b/kT\right) + \beta E^{1/2}\right] \quad (3)$$

$$\text{and, } \beta = \left(1/kT\right)\left(e^3/4\pi\varepsilon_0\varepsilon\right)^{1/2} \quad (4)$$

where $\alpha$ is a constant; $T$, absolute temperature; E, applied electric field; $\mu$, bulk mobility; $m_0$, electron mass; $m^*$, effective mass of electron; $\varphi_b$, Schottky barrier height (SBH); $k$, the Boltzmann constant; $\varepsilon_0$, the permittivity of free space; and $\varepsilon$, the optical dielectric constant.[34] This form of the Schottky equation is appropriate whenever the mean free path of the electrons is less than the Schottky barrier width, which is generally true in perovskite oxides, where it is typically about one unit cell. Simmons' equation has an 'extra factor' of $E$ outside the exponent, as compared to original Schottky equation.[33,34] Therefore, at low applied voltages, when exponential becomes approximately unity, the current density would shows its linear dependence on applied electric field and hence J-E curves looks like Ohmic.[4,32-35] We used Equation 3 to fit our J-E data corresponding to OFF-states in two parts: (i) in the low-voltage regime (< 0.5 V); and (ii) in the high-voltage regime (> 0.5 V). Figure 4(b) shows the OFF-state J-E characteristics corresponding to all four RS modes and can be well fitted by Eq. 3. The inset of Fig. 4(b) shows the straight line fit of J-E data in the low-voltage regime, confirming Simmons' modified Schottky conduction mechanism.

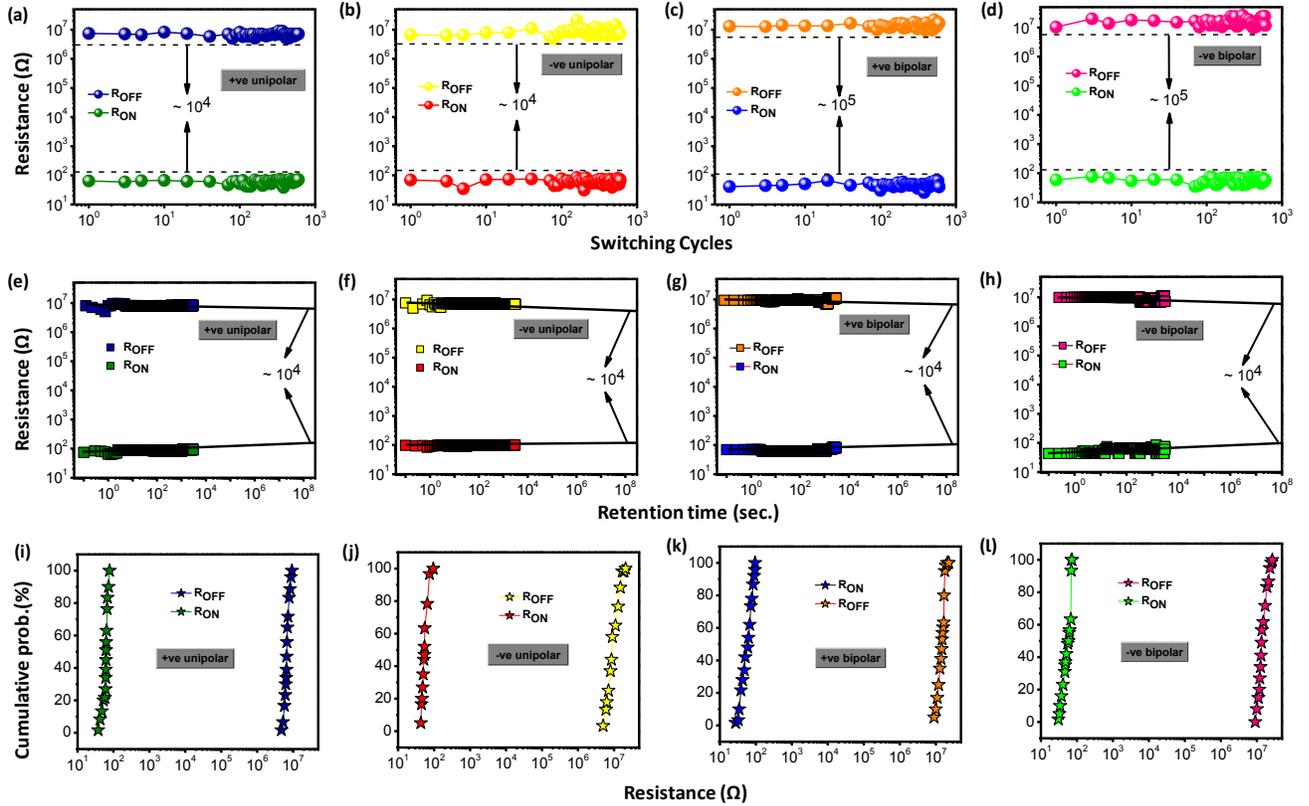

**Figure 5.** For all four RS modes of Pt/a-LHO/Pt memory cell: (a)-(d) the endurance characteristics; (e)-(h) retention performances at 370 K with power law extrapolation up to 10 years showing stable resistance values of ON and OFF-states; (i)-(l) uniform statistical distributions of ON and OFF resistance levels with mean values of around 35-100Ω and 5-12 MΩ, respectively

To confirm reliability and proper functionality, we investigated both endurance and retention characteristics of the Pt/a-LHO/Pt memory cell. Figures 5(a)-(d) show the endurance characteristics of the memory cell in all RS modes under a fixed $I_{CC}$ of 1 mA for up to 650 consecutive switching cycles in each mode. For all RS modes external applied voltages to set (ON) and reset (OFF) the memory cell was observed to show small voltage variation of about ± 5-8 %, with $V_{OFF}$ ~ ± 1.3-1.6 V and $V_{ON}$ ~ ± 3.6-4.2 V. The retention performance of all RS modes recorded at 370 K (ITRS specified higher operation temperature limit) at a read voltage of 0.1 V are shown in Fig. 5(e)-(h). The resistance ratio ($R_{OFF}/R_{ON}$) between the two resistance states for all these modes was found to be in the range of ~$10^4$-$10^5$ and remained unchanged over a time period of ~$10^3$ seconds. Further, the power law extrapolation suggests that the $R_{OFF}/R_{ON}$ ratio even after 10 years can still be expected to be ~$10^4$, which indicate excellent stability and retention of Pt/a-LHO/Pt memory cell.[36] Moreover, the statistical distributions of each RS modes were depicted in Fig. 5(i)-(l). The distributions of ON and OFF states were found to be quite uniform with mean values of around 35-100Ω and 5-12 MΩ, respectively. These observations suggest repeatable nonvolatile RS behavior of Pt/a-LHO/Pt memory cell corresponding to its different switching modes.

**IV. Conclusions**

In conclusion, nonpolar resistive switching phenomenon has been identified in Pt/a-LHO/Pt memory cell. XPS characterization along with temperature dependent switching studies revealed that formation (set) and rupture (reset) of mixed conducting filaments, formed out of oxygen vacancies and metallic Ho atoms, are responsible for the change in the resistance states of the memory cell. For all observed RS modes, the repeatability of set/reset processes in the memory cell was attributed to a Joule–heating assisted thermo-chemical effect. The reliability of the Pt/a-LHO/Pt memory cell was confirmed by the excellent endurance and retention performances. Our results demonstrated promising potential of ternary rare-earth dielectric LHO for non-volatile RRAM device applications.

**Acknowledgement:** Financial support from DOE Grant No. DE-FG02-08ER46526 is acknowledged. Y. S. is grateful to IFN for graduate fellowship under NSF-RII-0701525 grant.

S.P.P. thanks NSF for financial assistance under Grant No: NSF-EFRI RESTOR # 1038272. The authors are thankful to Mr. Oscar Resto for FE-SEM measurements.

**References:**

[1] International Technology Roadmap for Semiconductors (Semiconductor Industry Association, San Jose, CA, **2013**; see http:/www.itrs.net for updates.).

[2] D. S. Jeong, R. Thomas, R. S. Katiyar, J. F. Scott, H. Kohlstedt, A. Petraru, and C. S. Hwang, Emerging memories: resistive switching mechanisms and current status. *Rep. Prog. Phys.* **75**, 076502 (2012).

[3] R. Waser and M. Aono, Nat. Mater. **6**, 833-840 (2007).

[4] T. You, N. Du, S. Slesazeck, T. Mikolajick, G. Li, D. Burger, I. Skorupa, H. Stocker, B. Abendroth, A. Beyer, K. Volz, O. G. Schmidt, and H. Schmidt, ACS Appl. Mater. Interfaces **6**, 19758-19765 (2014).

[5] Y. Cui, H. Peng, S. Wu, R. Wang, and T. Wu, ACS Appl. Mater. Interfaces **5**, 1213-1217 (2013).

[6] D. Pantel, S. Goetze, D. Hesse, and M. Alexe, ACS Nano **5**, 6032-6038 (2011).

[7] C. S. Moreno, C. Munuera, S. Valencia, F. Kronast, X. Obradors, and C. Ocal, Nano Lett. **10**, 3828-3835 (2010).

[8] H. Lee, H. Kim, T. N. Van, D.-W. Kim, and J. Y. Park, ACS Appl. Mater. Interfaces **5**, 11668-11672 (2013).

[9] U. Celano, L. Goux, A. Belmonte, K. Opsomer, A. Franquet, A. Schulze, C. Detavernier, O. Richard, H. Bender, M. Jurczak, and W. Vandervorst, Nano Lett. **14**, 2401-2406 (2011).

[10] C.-W. Hsu and L.-J. Chou, Nano Letters **12**, 4247-4253 (2012).

[11] A. Younis, D. Chu, I. Mihail, and S. Li, ACS Appl. Mater. Interfaces **5**, 9429-9434 (2013).

[12] H. Tian, H.-Y. Chen, T.-L. Ren, C. Li, Q.-T. Xue, M. A. Mohammad, C. Wu, Y. Yang, and H. S. P. Wong, Nano Lett. **14**, 3214-3219 (2014).

[13] L. Chun-Chieh, C.-Y. Lin, L. Meng-Han, L. Chen-Hsi, and T.-Y. Tseng, Electron Devices, IEEE Trans. Electron Devices **54**, 3146-3151 (2007).

[14] L. Zhong, L. Jiang, R. Huang, and C. H. de Groot, Appl. Phys. Lett. **104**, 093507 (2014).

[15] H.-H. Huang, W.-C. Shih, and C.-H. Lai, Appl. Phys. Lett. **96**, 193505 (2010).

[16] S. P. Pavunny, R. Thomas, N. M. Murari, J. Schubert, V. Niessen, R. Luptak, T. S. Kalkur, and R. S. Katiyar, Integr. Ferroelectr. **125**, 44-52 (2011).

[17] L. Xu Gao, J. Yin, Y. Xia, K. Yin, L. Gao, H. Guo, and J. Zhiguo, J. Phys. D: Appl. Phys. **41**, 235105 (2008).

[18] J. M. J. Lopes, M. Roeckerath, T. Heeg, E. Rije, J. Schubert, S. Mantl, V. V. Afanas'ev, S. Shamuilia, A. Stesmans, Y. Jia, and D. G. Schlom, Appl. Phys. Lett. **89**, 222902 (2006).

[19] W. C. Chin, K. Y. Cheong, and Z. Hassan, Mater. Sci. Semicond. Proc. **13**, 303-314 (2010).

[20] P. Misra, Y. Sharma, and R. S. Katiyar, ECS Trans. **61**, 133-138 (2014).

[21] P. Misra, S. P. Pavunny, Y. Sharma, and R. S. Katiyar, Integr. Ferroelectr. **157**, 47-56 (2014).
[22] K. Li, Y. Xia, B. Xu, X. Gao, H. Guo, Z. Liu, and J. Yin, Appl. Phys. Lett. **96**, 182904 (2010).
[23] Y. Sharma, P. Misra, S. P. Pavunny, and R. S. Katiyar, Appl. Phys. Lett. **104**, 073501 (2014).
[24] S. Coh, T. Heeg, J. H. Haeni, M. D. Biegalski, J. Lettieri, L. F. Edge, K. E. O'Brien, M. Bernhagen, P. Reiche, R. Uecker, S. Trolier-McKinstry, D. G. Schlom, and D. Vanderbilt, Phys. Rev. B **82**, 064101 (2010).
[25] S. P. Pavunny, A. Kumar, P. Misra, J. F. Scott, and R. S. Katiyar, phys. status solidi (b) **251**, 131-139 (2013).
[26] Y. Sharma, S. Sahoo, A. K. Mishra, P. Misra, S. P. Pavunny, A. Dwivedi, S. M. Sharma, and R. S. Katiyar, J. Appl. Phys. **117**, 094101 (2015).
[27] J. P. Strachan, M. D. Pickett, J. J. Yang, S. Aloni, A. L. David Kilcoyne, G. Medeiros-Ribeiro, and R. Stanley Williams, Adv. Mater. **22**, 3573-3577 (2010).
[28] K. Szot, W. Speier, G. Bihlmayer, and R. Waser, Nat. Mater. **5**, 312-320 (2006).
[29] P. Tung-Ming, L. Chih-Hung, S. Mondal, and K. Fu-Hsiang, IEEE Trans. Nanotechnol. **11**, 1040-1046 (2012).
[30] P. Zhou, L. Ye, Q. Q. Sun, P. F. Wang, A. Q. Jiang, S. J. Ding, and D. W. Zhang, Nanoscale Res. Lett. **8**, 91 (2013).
[31] R. Waser, R. Dittmann, G. Staikov, and K. Szot, Adv. Mater. **21**, 2632-2663 (2009).
[32] J. G. Simmons, J. Appl. Phys. **34**, 1793-1803 (1963).
[33] J. F. Scott, J. Phys.: Condens. Matter **26**, 142202 (2014).
[34] S. Zafar, R. E. Jones, P. Chu, B. White, B. Jiang, D. Taylor, P. Zurcher, and S. Gillepsie, Appl. Phys. Lett. **72**, 2820-2822 (1998).
[35] J. G. Simmons, J. Appl. Phys. **35**, 2472-2481 (1964).
[36] I. Valov, R. Waser, R. J. Jameson, and M. N. Kozicki, Nanotechnology **22**, 254003 (2011).